\documentclass[12pt,preprint]{aastex}

\begin{document}

\title{Neutral Hydrogen Mapping of Virgo Cluster Blue Compact Dwarf Galaxies}

\author{G. Lyle Hoffman}
\affil{Dept. of Physics, Lafayette College, Easton, PA  18042; 
hoffmang@lafayette.edu}

\author{Noah Brosch}
\affil{Wise Observatory, Tel Aviv University, Tel Aviv 69978, Israel;
noah@wise.tau.ac.il}

\author{E.E. Salpeter}
\affil{Center for Radiophysics and Space Research, Cornell University, Ithaca, 
NY  14853; ees12@cornell.edu}

\and

\author{Nathan J. Carle}
\affil{Dept. of Physics, Lafayette College, Easton, PA  18042}

\received{}
\accepted{}
\slugcomment{To appear in AJ Dec. 2003}

\begin{abstract}

A new installment of neutral hydrogen mappings of Blue Compact Dwarf 
galaxies, as defined by optical morphology,
in and near the Virgo cluster is presented.
The primary motivation was to search for outlying clouds of \ion{H}{1}
as potential interactive triggers of the enhanced star formation, and
therefore the mapped galaxies were selected for large \ion{H}{1} mass, large 
optical diameter, and large velocity profile width.
Approximately half the sample proved to have one or more small, low column
density star-free companion clouds, either detached or appearing as an 
appendage in our maps, at resolution of order 4 kpc.
Comparison is made to a sample of similarly mapped field BCD galaxies
drawn from the literature; however, the Virgo cluster sample of mapped
BCDs is still too small for conclusive comparisons to be made.

We found, on the one hand, little or no evidence for ram pressure stripping 
nor, on the other, for extremely extended low column density \ion{H}{1} envelopes.
The \ion{H}{1} rotation curves in most cases rise approximately linearly,
and slowly, as far out as we can trace the gas.

\end{abstract}


\section{Introduction}

Although there are examples of Blue Compact Dwarf (BCD) galaxies which appear 
optically to consist of a current starburst with no underlying population of 
older stars \citep[e.g.]{TIL97,BAH98,HT95}, the majority of BCDs  consist of 
an intense starburst embedded in a low surface brightness (LSB)  envelope of 
previous generation stars \citep{KMV88,PLTF96,DDB97,AB98}.
For a general review of BCDs and the possibly related \ion{H}{2} galaxies, 
see \citet{KO00}.
For our purposes, the term ``BCD'' refers to galaxy morphology regardless
of galaxy size or luminosity, in the spirit of the morphological
classification in the Virgo Cluster Catalog (VCC) of \citet{BST85}.
The collection of BCDs in the VCC, even after objects with later-determined
redshifts that place them in the background of the cluster are removed,
is admittedly a heterogeneous sample; however, it is useful to consider
the set as a whole before finer distinctions are drawn among the objects.

One important question is how the size of the optically luminous region 
compares to the \ion{H}{1} envelope; what fraction of the \ion{H}{1} reservoir 
is involved in the starbursts, present or past?
The \ion{H}{1} profile widths from single beam Arecibo data for the
VCC BCDs are consistent 
with those for Sm and Im galaxies of the same total luminosity, but the 
optical diameters of the BCDs are smaller by a factor of 2 or 3 \citep{HHSL89}.
We would like to know if the \ion{H}{1} is similarly concentrated by 
comparison to other irregular galaxies of similar total mass. 
In particular, prior observations have not told us what fraction of BCDs 
have \ion{H}{1} 
extents greatly exceeding the diameter of the region that has formed stars 
[pertinent to the question of whether or not \ion{H}{1} envelopes of BCDs contribute 
significantly to the Lyman Limit Systems  (LLS) seen in absorption in QSO 
spectra]; in those cases where the \ion{H}{1} does exceed the optical extent 
by more than a factor of 2 it is still not clear whether the outlying gas is 
primordial and still infalling, distended by a blow-out from the central 
star-burst, or due to tidal interactions or ram-pressure sweeping.

\ion{H}{2} galaxies are 
identified by pronounced optical wavelength emission lines similar to those
from Galactic \ion{H}{2} regions while BCDs are identified by their morphology 
in photographic images (and we exclude from consideration those BCDs which were
subsequently found to have redshifts that place them far in the background 
of Virgo).
Identification solely by emission lines also produces a heterogeneous set of
objects, and  
one point of inquiry is to what extent the two sets (BCDs and \ion{H}{2}
galaxies) overlap.
Are external interactions responsible for the starbursts, either in the form 
of infalling \ion{H}{1} clouds or tidal interactions with other nearby galaxies?
The latter question has been addressed by \ion{H}{1} mapping with various 
resolutions, mainly for field BCDs and \ion{H}{2} galaxies 
\citep[e.g.]{vZSS98,BAH98,TBS93,TBGS95,
TBGS96}, with largely inconclusive results.
The BCDs in and around the Virgo cluster are particularly interesting in this 
regard, since they are distributed among the several groups and subclusters,
presumably at different stages of interaction with the rest of the cluster.
Are there subtle tidal effects due to the neighboring galaxies or due to the 
group potential, even though no evidence of dependence of star formation rates 
on Virgocentric distance can be found \citep{AB98}?
An effective strategy for assessing the prevalence of this phenomenon is 
to use Arecibo Observatory
\footnote{The Arecibo Observatory is part of the National Astronomy and 
Ionosphere Center, which is operated by Cornell University under a management 
agreement with the National Science Foundation.}
for cursory mapping of the surroundings of all VCC BCDs with redshifts in the 
range of the 
Virgo Cluster.  
Only 6 of the 35 VCC BCDs (with redshifts in the Virgo Cluster range) have 
mapping reported prior to this paper \citep{HSFRWH96,BAH98,LSvG87}. 
Here we add Arecibo mapping for another five, an installment toward a sample
sufficiently complete to answer the questions raised above. 

The objects found to have outlying emission in our Arecibo mapping are 
primarily those with the largest optical diameters, the largest single-beam 
\ion{H}{1} fluxes, and the largest \ion{H}{1} profile widths.
Consequently we undertook to extend the sample of VCC BCDs mapped at the 
Very Large Array
\footnote{The Very Large Array of the National Radio Astronomy Observatory 
is a facility of the National Science Foundation, operated under cooperative 
agreement by Associated Universities, Inc.} 
to include the 5 largest BCDs (along with one smaller object for which
Arecibo mapping indicated outlying gas and one BCD in Leo). 
We sought to integrate long enough to acquire detailed mapping of the outer regions of
the \ion{H}{1} disk down to a column density $N_{HI}$ of a 
few $\times 10^{19}$ ${\rm cm}^{-2}$.
The low spatial resolution mapping at Arecibo was designed to look for any very extended
\ion{H}{1} appendages down to even lower column density.
Ionization of hydrogen by extragalactic UV is thought to be important only for 
$N_{HI} < 2 \times 10^{19}\;{\rm cm}^{-2}$, and recent detections of 
``Mini-High Velocity Clouds'' with peak column densities well below 
$10^{19}$ ${\rm cm}^{-2}$ \citep{HSP02} may suggest even lower critical column densities
for ionization.
A significant contribution to low redshift LLS from BCDs would be detectable.
Our velocity range should have been adequate to 
detect any outlying \ion{H}{1} clouds (or \ion{H}{1}-rich, optically faint 
dwarf companions) associated with these galaxies, should they exist.

In Sect. 2 we give the details of the Arecibo mapping, and the VLA parameters
follow in Sect. 3.
The results from both sets of observations are presented in Sect. 4, along 
with a comparison of the maps for objects mapped at both observatories.
A few preliminary theoretical considerations are discussed in Sect. 5, followed
by conclusions and a summary in Sect. 6.

\section{Arecibo Observations}

Our mapping of the environs of five Virgo cluster BCDs was conducted as a 
commissioning phase project at the upgraded Arecibo Observatory
in July/August 1998.
We used the Gregorian feed system with the ``L narrow'' receiver in total 
power (position-switched) mode, with 6.10 kHz (about 1.3~km~${\rm s}^{-1}$) 
channel spacing.
Calibration was accomplished by observing several continuum sources from the 
VLA calibrator list, chosen to have small size compared to the $3\farcm2$ beam.
In addition, we reobserved several spiral galaxies for which we had high 
signal-to-noise pre-upgrade \ion{H}{1} measurements and which were known to 
be $\ll 3\farcm2$ in extent.

Each BCD was mapped with pointed ON/OFF observations at the optical center 
and in a hexagon spaced $3\farcm2$ away.
The orientation of each hexagon is given by the axis-crossings of the spectra 
in Figs. 1 and 2.
One object, Mk~1263 = FS~32 (i.e., number 32 in the list presented by 
\citet{FS90}), was also observed at positions twice as far out from the center.
For the most part, we observed each point in a single 5 min (ON) scan, but 
each point around VCC~0022 was observed for twice that amount of time.
Since these observations spanned sundown and the tie-down system at Arecibo 
had not yet been brought on-line, baselines were not always as smooth as users 
have come to expect for nighttime observations.
None of the mapped objects was extended enough for sidelobes to contribute 
significantly to the $3\farcm2$ ring, but the $6\farcm4$ points around 
Mk~1263 may be receiving some sidelobe emission from the center.

For each spectrum, we removed a low-order polynomial baseline and measured 
the flux integrated over the full width of the feature and the profile widths 
at 50\%, 80\% and 20\% of the peak of the spectrum.
The results for the central spectrum are reported in Table 1, along with the 
systemic heliocentric velocity $V_{sys}$ defined to be the midpoint between 
the 50\% points on opposite sides of the central spectrum.
We estimated a total flux for each galaxy by integrating spatially over the 
galaxy as detailed in \citet{HSFRWH96}.
The type listed in Table 1 is that recorded for each galaxy in the 
NASA/IPAC Extragalactic Database
\footnote{The NASA/IPAC Extragalactic Database is operated by the Jet Propulsion
Laboratory, California Institute of Technology, under contract with the National
Aeronautics and Space Administration.}
(NED).

\section{VLA D Array Mapping}

We selected as candidates for VLA mapping all VCC BCDs with 
\ion{H}{1} flux $> 2.0$ Jy-km ${\rm s}^{-1}$, 
optical diameter $D_{opt} >$ 0\farcm6, and single beam profile width 
$\Delta V_{50} > 100$ km ${\rm s}^{-1}$, a total of 5 objects.
Adding one smaller object and one object in the Leo group, both known from 
Arecibo mapping to be extended, brought us to a total sample of 7 BCDs for 
which we sought high dynamic range D array mapping.
In all, the VLA-mapped VCC BCDs are brighter in $B_T$ by 1.2 mag, larger 
in optical diameter by a factor of 1.7, and brighter in \ion{H}{1} flux
by a factor of 4.0 than the VCC BCDs that have not yet been mapped.
\ion{H}{1} spectral line mapping using 27 antennas in the D array of the 
Very Large Array
was conducted on 1999 May 28-30.
Observational details are given in Table 2.
Online Hanning smoothing was employed, and calibration was accomplished 
using sources 1219+285 (B1950) and 3C286 from the VLA calibrator list.
The data were calibrated and editted using standard tasks in the 
Astronomical Image Processing System (Classic AIPS).
Continuum subtraction was done in the $uv$ plane via UVBAS, and maps were 
made and CLEANed using IMAGR with zero-spacing fluxes estimated from our 
Arecibo map and with robustness set equal to 0.
After imaging, the data cube was corrected for the VLA primary beam.

The Mk~1263 field in fact contained four galaxies within the VLA primary beam: 
the BCD Mk~1263 itself, the Sm-ImIV pair CGCG~66-29 = FS~35 and FS~36, and 
(near the edge of the primary beam) the SA(s)c galaxy NGC~3389, all within 
the observed velocity range.
Results are presented for three of the four (FS~36 was not detected; see 
Sect. 4.2.1).
VCC~0010, 0024, 0172, 0459 and 1437 were all isolated within their primary beams, 
but the VCC~0340 field also contained VCC~0329, a 16.8 mag (B) ImV? galaxy, 
and VCC~0379, a very faint galaxy of uncertain morphology with no previous 
redshift measurement of any kind.
To our surprise, both VCC~0329 and VCC~0379 were detected in this field.

For each of the galaxies in the seven VLA fields, we display in Figs. 3-24 
a velocity-integrated \ion{H}{1} emission map (contours) superimposed on a 
greyscale optical image from the Digitized Sky Survey (hereafter DSS), 
a velocity field from 
the first moment of the cube, and a panel of contour maps of the individual 
channels bearing significant flux.
The spatially-integrated spectra are displayed in Fig. 25.

\section{Results}

\subsection{Arecibo Results}

We mapped 5 BCDs at Arecibo.
Significant emission was seen in beams $3\farcm2$ away from the center in 3 of 
those cases (Mk~1263, VCC~0024 and VCC~1437); in 2 cases (Mk~1263 and VCC~0024)
there is a clear velocity separation between points $3\farcm2$ from center on 
opposite sides of the galaxy, suggesting a rotating disk at least $2\farcm0$ 
in diameter.
In VCC~1437 we also detect emission on opposite sides of the galaxy, but the
Arecibo data alone does not give evidence of rotation.
We will discuss VCC~1437 further in Sect. 4.5.
Spectra from the several observed points around each galaxy are shown in 
Fig. 1 (for Mk~1263) and Fig. 2 (for the other 4 BCDs).
Four more Virgo BCDs have pre-upgrade Arecibo mapping (to lower sensitivity) 
by us \citep{HSFRWH96}; 2 of these (VCC~0340 and VCC~0459) gave signs of being 
extended to a significant fraction of the Arecibo beam and 2 did not 
(VCC~0010 and UGC~7354).
None of the objects appears disk-like on the POSS-II films.

\subsection{VLA Results}

All seven BCDs mapped at the VLA proved to be extended compared to the D array 
synthesized beam, with \ion{H}{1} envelopes that extend well beyond the 
regions where stars have formed.
The main disks of all of the objects exhibit solid body rotation, i.e., 
their rotation curves rise linearly as far as we can trace the gas, far outside
the region of high optical surface brightness.
Comparison to other (non-BCD) dwarf irregular galaxies is made in Sect. 5.3.
Three of the BCDs (Mk~1263, VCC~0172 and VCC~0459) have reasonably 
symmetric \ion{H}{1} disks, but four of the objects (detailed below) displayed 
appendages which 
might be tidal flares, gas blown out of the disks by multiple supernovae, or 
distinct infalling clouds as detailed below.
D array resolution is not adequate to allow us to distinguish which 
possibilities are more likely in any of the four cases.
Nor are we able to tell if the outer parts of these galaxies, aside from the 
apparent appendages, are irregular at more resolved angular scales.
Results for the detected galaxies are tabulated in Table 3, which has the 
following columns:  galaxy name, assumed distance, type as listed in NED, 
\ion{H}{1} flux integrated over velocity and the spatial extent of the galaxy,
the corresponding \ion{H}{1} mass, \ion{H}{1} diameter of the 
$10^{20}~{\rm cm}^{-2}$ contour on the integrated map in arcmin and kpc,
the gradient of the rotation curve averaged over the approximately linear 
portion of the rotation curve (for BCDs only), \ion{H}{1} mass to blue
luminosity ratio, and \ion{H}{1} to optical diameter ratio.

\subsubsection{The Mk~1263 field}

Within the primary beam field centered on Mk~1263, we also detected emission 
from the Sm galaxy CGCG~66-29 and the SA(s)c galaxy NGC~3389 (at the edge of 
the field).
Emission from the BCD Mk~1263 and from NGC~3389 appears disklike.
The \ion{H}{1} disk around Mk~1263 is misaligned by nearly $45\arcdeg~$ from 
the high surface brightness optical portion of the galaxy visible on the 
DSS (Fig. 3).
The rotation curve is approximately solid body in form.
The emission-bearing channel maps are displayed in Fig. 4.

The \ion{H}{1} disk around NGC~3389 is reasonably normal for a galaxy of its 
type; since the galaxy lies very near the edge of the primary beam field, we 
cannot determine whether the apparent appendage to the South of the \ion{H}{1} 
disk in Fig. 5 is real or due to noise.
Channel maps are presented in Fig. 6.

In \citet{HSFRWH96}, we attributed emission seen with the Arecibo beam to the 
South of CGCG~66-29 (called FS~35 in that paper) to the ImIV galaxy FS~36.
The VLA results shown in Fig. 5 forces us to alter that interpretation.
There is a cloud of \ion{H}{1} to the South of CGCG~66-29, but it is displaced 
about $3\arcmin~$ to the SW of FS~36 --- further from FS~36 than FS~36 is from 
CGCG~66-29.
There is a more irregular apparently detached \ion{H}{1} cloud a similar 
distance N of CGCG~66-29.
The two galaxies are $2\farcm7$ apart, at projected separation 19.9 kpc 
assuming a distance of 25.4 Mpc for the group (allowing for Virgocentric 
infall).
The nearest large galaxy is NGC~3389, $13\farcm9$ to the NW, at projected 
separation 103 kpc.
We are not aware of any intragroup medium having been detected in X-ray.
Some form of tidal interaction therefore seems the most likely explanation 
for the two detached \ion{H}{1} clouds, although it is difficult to envision 
a detailed tidal scenario which would completely strip FS~36 of its gas while 
also drawing out the northern tail (and perhaps the SW cloud as well) from 
CGCG~66-29.

\subsubsection{The Virgo BCDs}

IC~3017 = VCC~0010 has an appendage to the SW, perpendicular to the major axes 
of both the optical image and the main \ion{H}{1} disk, at velocities 
$\sim 70$~km~${\rm s}^{-1}$ below the systemic velocity.
The main \ion{H}{1} disk is reasonably well aligned with the optical image 
in Fig. 9 and displays quasi-solid-body rotation (rotation velocity rising
nearly linearly as far as we can trace the gas).
Channel maps are shown in Fig. 10.
An H band image is available in \citet{GBDFS03}.

VCC~0024 has a main \ion{H}{1} disk well aligned with the optical image as 
shown in Fig. 11, but at each end there is a flare of \ion{H}{1} kinked 
$20\arcdeg$ from the major axis, with a shallower scale length.
There is, in addition, an apparently separate, small \ion{H}{1} cloud with 
small velocity dispersion $2\farcm2$ (12 kpc) to the north, 
50~km~${\rm s}^{-1}$ below the systemic velocity of the galaxy.
It is not clear, from D array observations alone, whether the flares are 
simply warps in an extended disk, tidal flares, or distinct clouds.
The rotation curve is approximately solid-body in form, rising linearly
to the end and continues smoothly through the flares.
The small cloud to the north is detected at $\sim 5 \sigma$ and can be seen 
most clearly in the 1237~km~${\rm s}^{-1}$ panel of Fig.12, where there is 
a hint that there may be a low column density bridge to the main disk.
Images in B, V, and K are available in \citet{GBDFS03}.

VCC~0172 exhibits a relatively unremarkable \ion{H}{1} disk with no distinct 
appendages or detached clouds evident above our detection threshold, around 
$2 \times 10^{19}~{\rm cm}^{-2}$.
The rotation curve appears to flatten on the NW end of the disk but not 
(as much) on the SE end.
The integrated \ion{H}{1} map and velocity field are shown in Fig. 13, 
channel maps in Fig. 14.
\citet{GBDFS03} offer an image in H band.
\citet{BHA98} describe the H$\alpha$ morphology as ``C$+$E'' meaning
that there is one or more prominent star formation regions near the center
of the galaxy and one or more near the outer edge.

VCC~0340 has a small, apparently distinct cloud just off the end of the major 
axis of the \ion{H}{1} disk, $1\farcm3$ (7 kpc) N of the center of the galaxy.
The cloud has velocity 60~km~${\rm s}^{-1}$ higher than the galaxy's systemic 
velocity and very small velocity dispersion, and is unresolved by the D array.
At $5 \sigma$, it also begs confirmation.
The velocity field is complex, with some indications of tumbling about an EW 
axis but without a simple pattern.
An H band image is available in \citet{GBDFS03}.
The ImV dwarf VCC~0329 was detected near the edge of the primary beam field; 
our \ion{H}{1} map and velocity field are shown in Fig. 17, with channel maps 
in Fig. 18.
In the same primary beam field, there is a newly discovered (by us, in these 
D array observations) \ion{H}{1} cloud around VCC~0379, a 17.0 mag LSB typed ? 
by \citet{BST85}.
The details are shown in Figs. 19 and 20.

\subsubsection{VCC~1437 and a Possible Accreted Gas Cloud}

VCC~1437 = Mk 772 exhibits a disk in solid body rotation, but then has an obvious 
appendage to the NW at a velocity close to the 
systemic velocity, 20~km~${\rm s}^{-1}$ lower than the velocity of the 
W end of the main disk.
D array mapping suggests, but not conclusively, that this is a distinct cloud.
Prior C array mapping by \citet{LSvG87} was evidently not deep enough to 
detect the appendage.
Since the cloud is unresolved by the D array beam its diameter is $< 5$~kpc 
and it has a small velocity dispersion.
The integrated \ion{H}{1} mass is $4 \times 10^6$~${\rm M}_{\sun}$ assuming
a distance to VCC~1437 of 19 Mpc.
That makes the cloud very much comparable to the Compact High Velocity Clouds
(CHVC) observed around the Milky Way, if the CHVC have distances of order
150-200 kpc \citep{BBdH02}.
Images of VCC~1437 in H and H$\alpha$ are shown in \citet{GBDFS03}, and
\citet{VI03} contribute spectroscopy leading to chemical abundances for the
galaxy.
Those results, along with the significant \ion{H}{1} disk we have found,
render the E: type, recorded in the RC3 \citep{RC3} and repeated in NED,
highly suspect.

\subsection{Comparison of Arecibo and VLA Mapping}

In Table 4, results from the VLA mapping are compared with those from
Arecibo mapping for objects in common in the two samples.
The compared quantities are:
(1)  the diameter of the \ion{H}{1} envelope, measured to the outermost
point at which we can detect the gas with each instrument,
converted to kpc using the distance as given in Table 3;
(2)  the \ion{H}{1} velocity profile width (FWHM), $\Delta V_{50}$, in
km~${\rm s}^{-1}$; and
(3)  the integrated \ion{H}{1} mass $M_{HI}$ in units of 
$10^{8}$~${\rm M}_{\sun}$.
The final column gives the ratio of indicative dynamic mass $M_{ind} =
V_{rot}^{2} R / G$ to \ion{H}{1} mass, using radii measured from the VLA
maps, profile widths from Arecibo (since we have better velocity resolution
in that data) and \ion{H}{1} masses from Arecibo (since the single dish has
better sensitivity to outlying diffuse gas).
Since the optical images are too asymmetric to determine inclinations reliably, 
and the  velocity fields in the \ion{H}{1} maps are too irregular to allow us to
determine kinematical inclinations, we took $\sin^{2} i = 2 / 3$ for all BCDs.

In all cases save VCC~0172 (for which emission was found only in the central
beam at Arecibo) and VCC~0024, the Arecibo diameter measurement is larger
than the VLA measurement.
This is expected, since Arecibo is more sensitive to diffuse emission and
can reach lower column densities in the allotted time than the VLA.
However, the Arecibo diameters are not {\em much} larger, only by a few 
percent in most cases and never more than a factor of 2.
Similarly, the \ion{H}{1} masses are larger in the Arecibo maps
by only a few percent in most cases.
This suggests that the \ion{H}{1} envelopes are only a little larger than
shown in the VLA maps.
The Arecibo and VLA profile widths are comparable, within the resolution of
the VLA maps for the most part.
Since the rotation curves in the VLA maps mostly rise linearly as far as the
VLA can trace the gas, this comparision confirms that outlying gas does not
contribute greatly to the integrated profile. 

With the single exception of VCC~0340, all BCDs mapped with the VLA have 
$M_{ind}$ significantly larger than the total mass of gas plus stars could
be for any reasonable stellar mass-to-light ratio.
The W' cloud, of which VCC~0340 is a member, is a region of enhanced 
\ion{H}{1} deficiency \citep{SSSFM02}, so one possibility is that VCC~0340
has suffered more ram-pressure stripping than the other BCDs in our sample.
That would remove gas preferentially from the outskirts, reducing both
the measured \ion{H}{1} diameter and the velocity profile width, artifically
reducing $M_{ind}$.
The otherwise large values of $M_{ind} / M_{HI}$ suggest that these BCDs have
considerable dark matter halos.

\section{Discussion}

\subsection{Breakdown of Virgo BCDs by Cloud Membership}

The Virgo cluster is comprised of a number of subclusters, and superimposed 
on the cluster proper are a couple of clouds thought to lie at a distance 
larger by about a factor of two \citep{SSSGH02,GBSPB99,BPT93}.
These background clouds (M and W) contain a significant number (14) of the 
35 VCC BCDs, according to the membership assignments of \citet{BPT93}.
Of the 10 Virgo BCDs that have been mapped to date, 4 (VCC~0144, 0172, 0324 and 
0468) are members of the W cloud, one (VCC~0340) is a member of the W' cloud 
which is thought \citep{BPT93} to lie between the W cloud and the Southern 
Extension of the Virgo cluster, two (VCC~0010 and 0022) are members of the M cloud 
which is the other background cloud, one (VCC~1437) is a member of subcluster 
B (formerly the S' cloud, roughly centered on M49), one (VCC~0459) is a member 
of subcluster A (associated with M87), and VCC~0024 is a member only of the 
cluster at large, not of any distinct subcluster or cloud.
The W, M and W' clouds also have an enhanced ratio of BCDs to dwarf irregulars
(dI) as a whole (types Sdm, Sm, Im, BCD and uncertain, as assigned in the VCC):
12/28 (43\%), 2/4 (50\%) and 2/6 (33\%) respectively (16/38 or 42\% overall).
In contrast, the A and B subclusters along with galaxies assigned only to
the cluster at large have BCD to dI ratio 19/110, or 17\%.

The W and M cloud BCDs, as a group, have significantly smaller \ion{H}{1} 
angular diameters than the BCDs affiliated with subclusters A and B and the cluster 
at large.
Only one (VCC~0172) of the W$+$M cloud BCDs has an \ion{H}{1} diameter 
(at column density $\sim 10^{19}$~atoms~${\rm cm}^{-2}$) much in excess of 
$2\arcmin$ while 2 of the 3 BCDs affiliated with subclusters A and B or the 
cluster at large have extents well above $2\arcmin$.
The W' BCD has as small an extent as any of the BCDs, however.
Since we chose objects for mapping based on large \ion{H}{1} mass and large
optical diameter, definitive results must await mapping of a larger sample.
However, while there is no doubt considerable spread in the intrinsic 
\ion{H}{1} extents, these results are consistent with the W$+$M cloud BCDS 
being more distant by a factor of 2 as suggested by the Tully-Fisher 
relation applied to the larger members of those clouds \citep{SSSGH02, GBSPB99}.
We do not see any more nor less tendency toward \ion{H}{1} morphological 
peculiarities (e.g., appendages or spurs, outlying clouds, unusual velocity
fields) in the W$+$M cloud BCDs than in the nearer sample.

\subsection{Outlying \ion{H}{1} and Lyman Limit Systems}

Table 3 indicates that \ion{H}{1} diameters $D_{H}$ are larger than optical 
diameters $D_{25}$ in every measurable case, but not enormously larger.
The mean of the ratio $D_{H} / D_{25}$ for the 6 VCC BCDs is 3.29 with
a standard deviation of 1.59 (standard deviation of the mean 0.65) and
a median of 3.0.
The sample spans a range of 1.53 to 6.19 in the ratio.
By comparison, the 51 objects in our ``field'' sample from the literature 
(see Table 5) have a mean of 3.44 with standard deviation 2.37 (standard 
deviation of the mean 0.34) and a median of 2.5, spanning the range 0.5
to 9.3.
Admittedly, both the \ion{H}{1} and optical diameters for the field sample 
are less homogeneously measured.
Still, we have no evidence for any significant difference in this ratio.
This suggests that the Virgo BCDs--or this subsample at any rate--have not
undergone any significant ram pressure stripping, since that would have removed
some of the outermost gas.

BCDs are in general fairly rare in comparison to dI galaxies,
which have already been shown to have insufficient outlying gas to account
for the numbers of Lyman Limit Systems observed at higher redshift 
\citep{CSB01, DP01}.
Since $D_{H} / D_{25}$ is not {\em extremely} large for BCDs, these relatively
rare objects are not good candidates for higher redshift Lyman Limit Systems either.  

\subsection{Rotation Curves and Comparison with dI Galaxies}

The velocity fields for Mk~1263, IC~3017, and VCC~0459 have relatively evenly
spaced contours suggesting that the rotation curves of these galaxies rise 
linearly, in solid-body fashion, as far as we can trace the gas.
The outermost contours for VCC~0024 are much more widely spaced than those
inside, which might be taken as an indication that the rotation curve is
beginning to flatten.
However, the evident warp in the \ion{H}{1} map complicates the issue.
VCC~0340 is too poorly resolved and too complex for us to say much about
its rotation curve.
VCC~1437 is in solid-body rotation to the edge of the disk on both ends of
the major axis, but then falls abruptly back to the central velocity
on the northwest end.
We interpret this to be an infalling cloud, as discussed in Sect. 4.2.3
above, and not a turnover in the rotation curve of the disk itself.
On the southeast end the \ion{H}{1} velocity field map also suggests
a turn back toward the central velocity, but we have poor signal-to-noise
in that beam.
In all cases, however, we have only a few beams spanning the \ion{H}{1} disk.
Beam-smearing of these rotation curves must be significant.
It is unlikely, in particular, that we could identify a kink within the
central arcmin or so of any of the galaxies.
In spite of those complications, we have attempted to estimate an average
rotation curve gradient (up to an unknown factor of $\sin i$) over the
approximately linear portion of the rotation curve.
The values, listed in Table 3, span the range 
0.003-0.016~km~${\rm s}^{-1}{\rm pc}^{-1}$ with a median of 
0.008~km~${\rm s}^{-1}{\rm pc}^{-1}$.
By way of comparison, the H$\alpha$ velocity fields for field BCDs obtained by
\citet{PMCKD02} have generally steeper central gradients:  their range is 
0.027-0.543~km~${\rm s}^{-1}{\rm pc}^{-1}$ with a median of
0.080~km~${\rm s}^{-1}{\rm pc}^{-1}$, ten times larger than ours.
Whether the H$\alpha$ gradients reflect orderly rotation or highly non-circular
motions in the vicinity of the region of active star formation is unclear,
however.

References to \ion{H}{1} maps of field BCDs and related objects are collected 
in Table 5.
In many cases, the velocity fields are so irregular or the spatial resolution
so poor that the authors do not attempt to derive a rotation curve.
We have, however, attempted to characterize the published velocity fields
as to whether the rotation curve is linearly rising to the outermost point 
at which any semblance to regularity remains, or flattens out on one or
both sides of the major (kinematical) axis.
Of the 42 cases in which a velocity field is shown, 13 appear to be 
approximately solid body, 10 remain linear on one side of the major axis but
flatten on the other, 18 flatten on both sides, and one appears to be an ongoing
merger of 2 clouds.
Two general features are important for comparison with dIs and the discussion
in Sect. 5.4 which follows.
($i$)  While in most cases the angular resolution is not small compared to the
size of the region where star formation is important, the slow rise of the
linear portion of the rotation curve makes it clear that the rotation velocity
must be quite small over that region.
However, ($ii$) the rotation velocity rises to an appreciable value, of order
50~km~${\rm s}^{-1}$, in the well-resolved outermost regions of the \ion{H}{1} disk.
This is comparable to the maximum velocities reached in the optical velocity fields
of \citet{PMCKD02}.

Typical dI galaxies with \ion{H}{1} mapping reported in the 
literature mostly have rotation curves that rise linearly and somewhat rapidly
in the central regions, then kink to a shallower slope.
For example, in \citet{S99} rotation curves for 62 late-type dwarf galaxies 
are shown.
Of these, 51 have rotation curves that are kinked, clearly steeper in the 
central regions than in the outskirts of the disk.
Only eleven have rotation curves that appear linear as far as they can be 
traced.
Two general features for dI are:
($i$) star formation extends well beyond the kink in the rotation curve, 
i.e., to regions where the rotation velocity is already of order the maximum
value, and ($ii$) this maximum value is typically of the same order as for BCDs,
$\sim 50$~km~${\rm s}^{-1}$.

\subsection{Dark Matter Distribution and Instabilities}

For dI there is still a little controversy about the detailed shape of the
rotation curve in the innermost region \citep{MOCFCMZ02, SMBB03}, but there is 
agreement that this region is small compared to where most of the gas and
stars are found and that appreciable rotation velocities are reached in those
outer regions.
The dark matter mass thus greatly exceeds the baryon mass and, even if there were
strong starbursts in the central regions, dynamical instabilities are unlikely.

For BCDs one has to make a distinction between the inner and outer regions
when considering questions of ``blowup'' and ``blowout'' \citep{DS86, MF99}.
For the outer regions, and for the galaxy as a whole, the dark matter mass
is much larger than the gas mass, and starbursts are not likely to eject much
material from the galaxy as a whole.
For the inner region, where the stars are, the situation is less clear.
The stars are young, massive and luminous, but the total stellar mass is not
very large.
Since the rotation velocity is so small here, the total indicative gravitational
mass is only moderately large.
The resolution of our VLA D array \ion{H}{1} mapping is not sufficient to tell
whether the gas is almost as concentrated as the stars or almost uniform density
over a VLA beam.
If it is concentrated, the gas mass could exceed the sum of stellar and dark matter 
mass within the star-forming region, and a very energetic starburst could not only
eject most of the gas from the inner region, but also cause a distension of the
stellar and dark matter distribution.
This distension is purely dynamical, somewhat as in \citet{GCS83} and \citet{NEF96},
but the stars and dark matter are only lifted into somewhat larger orbits.
The gas, on the other hand, can fall back into the inner region after it cools.
Better angular resolution \ion{H}{1} mapping of the inner regions is clearly 
needed to investigate this conjecture.

\subsection{Statistics on Companion \ion{H}{1} Clouds}

Several studies seeking indications of external interactions with field 
BCDs (or objects that would resemble the VCC BCDs on comparable plate
material at a comparable distance) have been conducted
\citep[e.g.]{SG00,vZSS98,BAH98,TBS93,TBGS95,TBGS96}.
Similar efforts have been made to find potential interactors that might,
a few hundred My in the future, turn otherwise isolated dwarf irregular
galaxies into BCDs \citep{PWL02,SG00,PW99,TTBS96}.
In most cases in which a distinct companion is found, careful inspection of the 
DSS reveals that the companion is a Low Surface Brightness galaxy.
\citet{PKLU01}, in a search for {\em optical} companions in the vicinity 
of 86 BCDs from the Second Byurakan Survey, concluded that $\sim80\%$ of their
sample were plausibly triggered either by tidal interaction with a companion
(either larger or smaller than the BCD) or by a recent merger.
\citet{OABMBM01} and \citet{BO02} argue on the basis of H$\alpha$ kinematics
that the several blue compact galaxies in their sample result from mergers
of gas-rich dwarf galaxies or massive \ion{H}{1} clouds.
 Detached \ion{H}{1} clouds free of stars to the detection limit of the DSS,
however, appear to be quite rare.

Most \ion{H}{1} surveys, like the present one, have been conducted with 
the VLA in its lowest spatial resolution mode, the D array, and have 
synthesized beams not much smaller than the \ion{H}{1} size of the BCD.
Therefore, for galaxies at roughly the distance of the Virgo Cluster, a gas
cloud separated from the \ion{H}{1} envelope of the BCD by less than the
\ion{H}{1} radius of the BCD might appear to be connected to that envelope, 
much like the VCC~1437 cloud discussed above.
We have scoured the maps available in the literature (see Table 5) for
such potential clouds as well as clearly separated clouds around field 
objects that would resemble the VCC BCDs at a comparable distance on comparable
plate material.
In the vicinity of 51 field (or loose group) BCDs, 
we find 20 with no companion objects within $\sim 200$ kpc;
13 with companion \ion{H}{1} clouds (including previously catalogued
galaxies) that have stars evident on the DSS or
other optical images available to the authors of those maps;
12 with clearly separated, spatially unresolved, low column density 
\ion{H}{1} clouds with no stars visible on available optical images;
and 9 with appendages or spurs that do not appear to be tidal in nature
and which might be detached from the BCD's \ion{H}{1} envelope at
higher spatial resolution.
For our 6 Virgo BCDs, the corresponding numbers are 2, 1, 2 and 2.
(In both sets of BCDs, some BCDs appear in multiple categories.)
Our sample of Virgo BCDs is not yet large enough to say whether or not there 
are differences in the two distributions.

\section{Conclusions and summary}

We have presented Arecibo maps of five VCC BCD galaxies and VLA D array maps 
of six along with a field in Leo containing one BCD, a Sm-ImIV pair, and
a much larger spiral.
The VCC maps form an installment toward an eventual sample statistically
complete enough to contrast with the growing field sample from the literature.

Of the five BCDs in the Virgo cluster and surrounding clouds mapped at Arecibo, 
three gave evidence of \ion{H}{1}
emission extended outside the central $3\farcm2$ beam.
That increases the sample of VCC BCDs with Arecibo mapping to nine objects,
with five in all showing emission outside the central beam.
The extended objects were those with the largest \ion{H}{1} masses, largest
optical diameters, and largest \ion{H}{1} velocity profile widths, and so
we used those criteria to select the six VCC BCDs for mapping with the 
VLA D array.

All seven BCDs mapped at the VLA are clearly extended beyond the D array 
beam, $\sim 45\arcsec$, and all give some indications of systematic rotation.
We found two examples of kinematically distinct appendages which might resolve
into distinct gas clouds at higher resolution, and two clearly detached
low column density clouds (one appearing to be a Compact HVC in the process
of coallescing with the galaxy's disk).
In all cases the clouds are within $\sim 10$ kpc of the BCD.

Comparison was made to a sample of similarly mapped field BCDs (or related objects)
drawn from 
the literature, but conclusive statements were not possible due to the
small size of the Virgo sample mapped to date.
However, regarding the question of starburst triggering by interactions,
we note that about one-third of the BCDs in both samples discussed here do
not appear to have companion \ion{H}{1} clouds, with or without stars.
Mapping of these objects at higher angular resolution, and of additional Virgo objects,
is clearly indicated.
Determinations of the hot (ionized) gas content in BCDs would also be very
helpful in determining whether the kinematically distinct clouds are falling
in or being blown out.

We have one fairly firm double negative result:
There is little evidence for any appreciable mass loss due to ram pressure stripping,
but also no evidence for any extremely extended \ion{H}{1} envelope at column
densities $N_{HI} < 2 \times 10^{19}$~${\rm cm}^{-2}$.
BCD galaxies are therefore not good candidates for Lyman Limit Systems.

The \ion{H}{1} rotation curves rise linearly and slowly all the way out for
BCDs, with maximum velocity and total indicative mass comparable to those
for the more common low surface brightness dwarf irregular galaxies.
On the other hand, the angular momentum in the very small central star-forming
region is quite small.
Better \ion{H}{1} angular resolution will be required to find out just how small 
the angular momentum is and to see just how concentrated the gas is.

\acknowledgments
We thank the anonymous referee for comments which helped increase the 
clarity of this paper.
The Digitized Sky Surveys were produced at the Space Telescope Science 
Institute under U.S. Government grant NAG W-2166. 
The images of these surveys are based on photographic data obtained using 
the Oschin Schmidt Telescope, which is operated by the California Institute 
of Technology and Palomar Observatory on Palomar Mountain. 
The plates were processed into the present compressed digital form with the 
permission of that institution.
The Second Palomar Observatory Sky Survey (POSS-II) was made by the 
California Institute of Technology with funds from the National Science 
Foundation, the National Geographic Society, the Sloan Foundation, the 
Samuel Oschin Foundation, and the Eastman Kodak Corporation.

\clearpage

\begin{deluxetable}{c c c c c c c c c c c}
\tabletypesize{\footnotesize}
\rotate
\tablewidth{0pt}
\tableheadfrac{0.2}
\tablecaption{Arecibo Observations}
\tablecolumns{11}
\tablehead{
\colhead{Galaxy} & \colhead{RA(1950)} & \colhead{Dec(1950)} & \colhead{Type} &
\colhead{Central Flux} & \colhead{rms} & \colhead{Total Flux} & 
\colhead{$V_{sys}$} & \colhead{$\Delta V_{50}$} & 
\colhead{$\Delta V_{80}$} & \colhead{$\Delta V_{20}$} \\
\colhead{} & \colhead{hhmmss.s} & \colhead{ddmmss} & \colhead{} &
\colhead{mJy km ${\rm s}^{-1}$} & \colhead{mJy} & 
\colhead{mJy km ${\rm s}^{-1}$} & \colhead{km~${\rm s}^{-1}$} & 
\colhead{km~${\rm s}^{-1}$} & \colhead{km~${\rm s}^{-1}$} & 
\colhead{km~${\rm s}^{-1}$} }
\startdata
Mk~1263 & 104617.8 & 122716 & BCD & 5515 & 1.5 & 5600 & 1321 & 129.9 & 114.6 & 
143.7 \\
VCC~0022 & 120751.0 & 132654 & BCD? & 643 & 1.0 & 700 & 1695 & 51.5 & 33.5 & 69.0 \\
VCC~0024 & 120803.0 & 120218 & BCD & 3902 & 1.5 & 3800 & 1292 & 197.5 & 181.5 & 
214.2 \\
VCC~0468 & 121846.2 & 42118 & BCD? & 880 & 1.7 & 700 & 1980 & 35.4 & 21.1 & 58.5 \\
VCC~1437 & 123001.2 & 92654 & E: & 1746 & 1.4 & 2500 & 1148 & 71.0 & 51.1 & 101.8 \\
\enddata
\end{deluxetable}

\begin{deluxetable}{l l l l l l l l}
\tabletypesize{\footnotesize}
\rotate
\tablewidth{0pt}
\tablecaption{Very Large Array Observations}
\tablehead{
\colhead{} & \colhead{Mk~1263} & \colhead{VCC~0010} & \colhead{VCC~0024} & 
\colhead{VCC~0172} & \colhead{VCC~0340} & \colhead{VCC~0459} & \colhead{VCC~1437} }
\startdata
Date (all 1999 May) & 28 & 29 & 29 & 29-30 & 30 & 29 & 29-30 \\
Antennas & 27 & 27 & 27 & 27 & 24 & 27 & 27 \\
Pointing Center R.A.(2000) & 10:49:03.0 & 12:09:24.8 & 12:10:36.2 & 
12:16:01.0 & 12:19:21.8 & 12:21:11.5 & 12:32:33.5 \\
Pointing Center Dec.(2000) & 12:18:00 & 13:34:25 & 11:45:37 & 04:39:02 & 
05:54:51 & 17:38:16 & 09:10:25 \\
$V$ (heliocentric, km ${\rm s}^{-1}$) & 1340 & 1972 & 1289 & 2175 & 1560 & 
2099 & 1160 \\
Array & D & D & D & D & D & D & D \\
Channels & 64 & 64 & 64 & 64 & 64 & 64 & 64 \\
Channel separation (km ${\rm s}^{-1}$) & 10.4 & 10.4 & 10.4 & 10.4 & 10.4 & 
10.4 & 10.4 \\
Time on source (min) & 53 & 53 & 54 & 92 & 79 & 81 & 110 \\
Beam (arcsec) & $49 \times 46$ & $49 \times 46$ & $47 \times 44$ & 
$49 \times 44$ & $47 \times 42$ & $45 \times 44$ & $50 \times 46$  \\
rms (mJy ${\rm beam}^{-1} {\rm chan}^{-1}$) & 1.2 & 1.4 & 1.3 & 1.0 & 1.5 & 
0.9 & 0.9 \\
\enddata
\end{deluxetable}

\begin{deluxetable}{l c c c c c c c c c}
\tabletypesize{\footnotesize}
\rotate
\tablewidth{0pt}
\tableheadfrac{0.3}
\tablecaption{Very Large Array Results}
\tablehead{
\colhead{Galaxy} & \colhead{Type} & \colhead{Distance} & \colhead{HI Flux} & 
\colhead{HI Mass} & \multicolumn{2}{c}{HI Diameter} & \colhead{Vel. Grad.} &
\colhead{${M_{H}}/{L_{B}}$} & \colhead{${D_{H}}/{D_{25}}$} \\
\colhead{} & \colhead{} & \colhead{Mpc} & \colhead{Jy km ${\rm s}^{-1}$} & 
\colhead{$10^{8} M_{\sun}$} & \colhead{arcmin} & \colhead{kpc} & 
\colhead{km ${\rm s}^{-1}{\rm pc}^{-1}$} &
\colhead{${M_{\sun}}/{L_{\sun}}$} & \colhead{} }
\startdata
NGC~3389 & SA(s)c & 25.4 & 13.66 & 20.8 & 3.2 & 23.6 & \nodata & 0.11 & 1.16 \\
Mk~1263 & BCD & 25.4 & 4.69 & 7.14 & 2.5 & 18.5 & 0.008 & 1.46 & 3.61 \\
CGCG~66-29 & Sm & 25.4 & 2.93 & 4.48 & 1.4 & 10.4 & \nodata & 0.64 & 1.28 \\
IC~3017 & BCD & 32.3 & 1.91 & 4.68 & 1.6 & 14.9 & 0.014 & 0.33 & 2.59 \\
VCC~0024 & BCD & 19.0 & 2.50 & 2.13 & 3.4 & 18.8 & 0.016 & 0.87 & 6.19 \\
VCC~0172 & IAm? & 32.3 & 4.72 & 11.6 & 2.5 & 23.4 & 0.008 & 0.66 & 3.53 \\
VCC~0329 & I? & 26.6 & 0.87 & 1.45 & 0.4 & 3.0 & \nodata & 0.69 & 1.20 \\
VCC~0340 & \nodata & 26.6 & 1.84 & 3.08 & 1.3 & 10.3 & 0.003 & 0.16 & 1.53 \\
VCC~0379 & \nodata & 26.6 & 1.10 & 1.84 & 1.1 & 8.7 & \nodata & 1.05 & \nodata \\
VCC~0459 & BCD & 19.0 & 2.46 & 2.10 & 1.6 & 8.7 & 0.009 & 0.36 & 2.54 \\
VCC~1437 & E: & 19.0 & 1.50 & 1.28 & 1.4 & 7.9 & 0.003 & 0.43 & 3.36 \\
\enddata
\end{deluxetable}

\begin{deluxetable}{l r r r r r r r}
\tabletypesize{\footnotesize}
\rotate
\tablewidth{0pt}
\tableheadfrac{0.3}
\tablecaption{Comparison of Arecibo and VLA Results}
\tablehead{
\colhead{Galaxy} & \multicolumn{2}{c}{$D_{HI}$ (kpc)} & 
\multicolumn{2}{c}{$\Delta V_{50}$ (km ${\rm s}^{-1}$)} & 
\multicolumn{2}{c}{$M_{HI}$ ($10^8 M_{\sun}$)} & \colhead{$M_{ind} / M_{HI}$} \\
\colhead{Name} & \colhead{VLA} & \colhead{AO} & 
\colhead{VLA} & \colhead{AO} & \colhead{VLA} & \colhead{AO} & \colhead{} }
\startdata
Mk~1263 & 18.5 & $\sim 22$ & 120 & 130 & 7.1 & 8.5 & 16.0 \\
IC~3017 & 14.9 & $< 28$ & 190 & 170 & 4.7 & 5.9 & 31.8 \\
VCC~0024 & 18.8 & $\sim 17$ & 190 & 198 & 2.1 & 3.3 & 97.4 \\
VCC~0172 & 23.4 & \nodata & 120 & 113 & 11.6 & 11.8 & 11.0 \\
VCC~0340 & 10.3 & $\sim 15$ & 60 & 49 & 3.1 & 6.0 & 1.8 \\
VCC~0459 & 8.7 & $\sim 11$ & 100 & 110 & 2.1 & 2.0 & 22.9 \\
VCC~1437 & 7.9 & $\sim 17$ & 50 & 71 & 1.3 & 2.1 & 8.3 \\
\enddata
\end{deluxetable}

\begin{deluxetable}{l c l p{3in} r}
\tabletypesize{\footnotesize}
\rotate
\tablewidth{0pt}
\tablecaption{Mapped BCDs Outside of Virgo}
\tablehead{
\colhead{Observers} & \colhead{Date Mapped} & \colhead{Telescope} & 
\colhead{Galaxies} & \colhead{Ref} \\}
\startdata
Brinks \& Klein & 1985 & VLA B & II~Zw~40 & 1 \\
Cox et al. & 1994\&1995 & VLA C\&D & II~Zw~70 & 2 \\
Hoffman et al. & 1988\&1989 & NAIC & UGC~7257, BTS~171 & 3 \\
Kobulnicky \& Skillman & 1993 & VLA DnC & NGC~5253 & 4 \\
Meurer et al. & 1992\&1993 & ATCA & NGC~2915 & 5 \\
Meurer, Staveley-Smith \& Killeen & 1990\&1991 & ATCA & NGC~1705 & 6 \\
Pustilnik et al. & 1994\&1995 & VLA C\&D & SBS~0335--052 & 7 \\
Putman et al. & 1996\&1997 & ATCA & FCC~35 & 8 \\
Simpson \& Gottesman & 1993 & VLA C & 
A1116+51; Haro~4, 27, 33, 36; Mrk~51, 67, 328 & 9 \\
Stil \& Israel & 1989\&1990 & WSRT & NGC~1569 & 10 \\
Taylor et al. & 1991\&1992 & VLA C\&D & 
VII~Zw~8, Mrk~314, 600, Haro~26 & 11, 12 \\
Taylor et al. & 1992\&1994 & VLA D & 
UM~323, 372, 422, 439, 446, 452, 456, 461/2, 463, 465, 483, 491, 
500/1, 504, 533, 538, 559 & 13 \\
van Zee et al. & 1993\&1994 & VLA C\&D & UGCA~20 & 14 \\
van Zee, Salzer \& Skillman & 1998-2000 & VLA B\&CS & 
Mrk~324, 750, 900, 1418; UM~38, 323 & 15 \\
van Zee, Skillman \& Salzer & 1997 & VLA B \& C & 
II~Zw~40, UGC~4483, UM~439, 461/462 & 16 \\
van Zee et al. & 1993\&1995 & VLA B\&C\&D & I~Zw~18 & 17 \\
Viallefond \& Thuan & 1979 & WSRT & I~Zw~36 & 18 \\
Walter et al. & 1986\&1990 & VLA B\&C & II~Zw~33 & 19 \\
\enddata
\tablerefs{(1) \citet{BK88}; (2) \citet{CSWvM01}; (3) \citet{HSFRWH96}; 
(4) \citet{KS95}; (5) \citet{MCBF96}; (6) \citet{MSK98}; (7) \citet{PBTLI01}; 
(8) \citet{PBMSSF98}; (9) \citet{SG00}; (10) \citet{SI98}; 
(11) \citet{TBPS94}; (12) \citet{TBS93}; (13) \citet{TBGS95}; 
(14) \citet{vZHSB96}; (15) \citet{vZSS01}; (16) \citet{vZSS98}; 
(17) \citet{vZWHS98}; (18) \citet{VT83}; (19) \citet{WBDK97}}
\end{deluxetable}

\end{document}